\documentclass[aps,twocolumn]{revtex4}
\usepackage{epsfig,amsmath,verbatim}

\newcommand {\vf}[1]{{\mathbf{#1}}}     %vector face
\newcommand {\vfq}{\vf{q}}      %vector q
\newcommand {\gc}{g^{(2)}}
\newcommand {\cpfree}{$c_p^{\text{free}}$}

\begin{document}

\title{\bf Glasses in hard spheres with short-range attraction}

\author{K.N.~Pham, S.U.~Egelhaaf, P.N.~Pusey, W.C.K.~Poon}

\affiliation{School of Physics, The University of Edinburgh, Mayfield
Road, Edinburgh EH9 3JZ, U.K.}

\date{\today}

\begin{abstract}
We report a detailed experimental study of the structure and dynamics
of glassy states in hard spheres with short-range attraction. The
system is a suspension of nearly-hard-sphere colloidal particles and
non-adsorbing linear polymer which induces a depletion attraction
between the particles. Observation of crystallization reveals a
re-entrant glass transition. Static light scattering shows a
continuous change in the static structure factors upon increasing
attraction. Dynamic light scattering results, which cover 11 orders of
magnitude in time, are consistent with the existence of two distinct
kinds of glasses, those dominated by inter-particle repulsion and
caging, and those dominated by attraction. Samples close to the `A3
point' predicted by mode coupling theory for such systems show very
slow, logarithmic dynamics.

\end{abstract}

\maketitle

\section{Introduction}\label{s:Intro}

Glassy states are found in many systems \cite{glassReview1}. However,
understanding the glass transition is still a major challenge for
statistical and condensed-matter physics. Simple and
well-characterized models hold an important place in this field. The
glass transition in the simplest model colloid, a suspension of hard
spheres, has been studied in detail for more than a decade
\cite{HSNature,HSglass,MCTtheoryHS,MCTreview}. The phase behaviour of
a system of $N$ hard-spheres of radius $R$ in volume $V$ is determined
by a single variable, the density or volume fraction $\phi=(4/3)\pi
R^3N/V$. Increasing $\phi$ drives the system from a stable fluid to a
fluid-crystal coexistence, and then a fully crystallized phase
\cite{HSNature}, which should be the thermodynamically favorable phase
up to $\phi = \pi/3\sqrt{2} \approx 0.74$. However, at $\phi\geq
\phi_g\approx0.58$, hard spheres fail to crystallize
\cite{HSNature,HSglass}. This is usually interpreted as a glass
transition due to the caging of particles by each other. The most
successful theoretical account given of this transition to date is
from mode coupling theory (MCT) \cite{MCTtheoryHS,HSglass}. Within
this framework, the coupling between different density fluctuation
modes drives the system into a dynamically arrested state
\cite{MCTreview}.

More recently, the focus of attention has moved on to hard spheres
with a short-range attraction. Besides being a good model for
understanding the fundamentals of the glass transition, such `sticky
hard spheres' are also ubiquitous in applications. Attraction in
hard-spheres can be realized experimentally by adding non-adsorbing
polymers to colloids. The center of mass of a polymer coil of radius
of gyration $r_g$ is excluded from a zone of width $\sim r_g$ from the
surface of each colloid. When two colloids come close enough to each
other so that their polymer-excluded regions overlap, the imbalance in
polymer osmotic pressure pushes them together. This effective
`depletion' attraction is well described \cite{depletion} by the
Asakura-Oosawa form \cite{AO}. Its dimensionless range can be
estimated by the ratio $\xi=r_g/R$, while its strength is governed by
the concentration of polymer coils in the free volume available to
them, \cpfree. The free volume depends on the exact structure of the
suspension, and is not an easily obtained experimental
parameter. However, the concentration of polymer in the whole system,
$c_p$, can be used as an alternative variable to describe the
composition of the system \cite{cptheory}.

The presence of a short-range attraction in hard spheres widens the
equilibrium fluid-crystal coexistence region in the phase diagram
\cite{cptheory,cpphaseexpt}, and introduces (non-equilibrium) gels
at low volume fractions \cite{lowPhiCP1,lowPhiCP2} and a re-entrant
glass transition at high volume fractions \cite{gelMCTYukawa,
gelGlassYukawa,glassGlassLogDecay,glassMicellar1,glassMCT,reentrantGlass,
reentrantSim,glassMicellar2,gelMCTHiPhi,SciortinoLogtime,
SciortinoAging}. In this paper we report a comprehensive study of
structure and dynamics in the vicinity of this re-entrant glass
transition in a model colloid-polymer mixture: sterically-stabilized
polymethylmethacrylate particles with added linear polystyrene
\cite{poon_review}.

We used light scattering to study the structure of colloids by
measuring the static structure factor (SSF), $S(q)$, which is
effectively the Fourier transform of the pair correlation
function. Dynamic light scattering was used to obtain the normalized
collective dynamic structure factor (DSF), $f(q,\tau)$, which measures
the time correlation of particle density fluctuations at wave vector
$q$ after delay time $\tau$. Our results agree in broad outline with
previous experimental studies and the trends predicted by MCT, while
the detailed nature of our study and the very wide time window of our
measured DSFs (11 orders of magnitude) together shed new light on the
nature of the re-entrant glass transition in sticky hard spheres.
Preliminary reports of some of these data have been given before
\cite{reentrantGlass,unsticking}.

\section{Experimental methods}

\subsection{Sample preparation}

The colloidal particles used in this study were
poly-methylmethacrylate (PMMA) spheres sterically stabilized by
chemically-grafted poly-(12-hydroxystearic acid) (PHSA) dispersed in
{\it cis}-decalin \cite{PMMAmaking}. The solvated PHSA, approximately 10nm
thick, produces a nearly-hard-sphere interaction between the colloidal
particles \cite{auer}. The particle radius, $R=202$nm, was determined
from the lattice spacing of the crystal phase at fluid-crystal
coexistence, taking the crystal to be at $\phi =0.545$. Particle size
polydispersity was 0.069, measured from the apparent angle dependence
of the diffusion coefficient in a dilute suspension
\cite{measurePolydisp}.

The colloidal volume fraction was calibrated by measuring the amount
of crystal phase in the coexistence region and taking the fluid and
crystal volume fraction to be at 0.494 and 0.545 respectively. Samples
of the volume-fraction-calibrated stock were also weighed and dried in
a vacuum oven to determine the effective density of the particles,
which was then used in subsequent calculations of the volume fraction
of samples prepared from the stock. The stock volume fraction was also
re-calibrated from time to time by drying and weighing.

To induce attraction between the colloids, we added non-adsorbing
linear polystyrene. This well-characterized model colloid-polymer
mixture has been studied extensively over the last decade
\cite{poon_review}. The polymer used was purchased from Polymer
Laboratories and had a molecular weight of $M_w=370000$ daltons. Its
radius of gyration in {\it cis}-decalin at $20^{\circ}$C was
calculated from the data in \cite{PSdata} to be $r_g = 17.8$~nm. Thus
the dimensionless range of the depletion attraction is $\xi \sim
r_g/R=0.09$.

Colloid-polymer mixture samples were prepared by mixing stocks of
colloids and polymers with known concentration and solvent by
weight. Sample volumes were about 1 cm$^3$. Knowing the density of
each species permits calculation of the final composition.

The main uncertainty in sample composition comes from a systematic
uncertainty in the calibrated volume fraction of the colloid
stock. This is because the volume fractions of coexisting fluid and
crystal phases for slightly polydisperse hard-spheres are slightly
different from those in a monodisperse colloid, but the exact values
are uncertain
\cite{polyFractionateBolhuis,polyFractionateBartlett}. However, all
samples were prepared from the same stocks of colloids and polymer
solutions, or stocks calibrated against each other. Some samples were
also derived from others in a controlled way (see next
paragraph). Therefore despite some systematic uncertainties in the
estimation of absolute volume fractions due to polydispersity, the
uncertainties in sample compositions relative to each other were
mostly from random errors in weighing. These uncertainties are below a
percent in the worst case and are insignificant in this work.

Samples were tumbled for prolonged periods of time to ensure proper
mixing of the components. After homogenizing, a small amount of each
sample was transferred to 3~mm-inner diameter glass tubes and sealed
for light scattering experiments. The rest of the sample was then left
undisturbed for visual observation of any phase transitions until
sedimentation appeared. Then some samples may be diluted with solvent
while others were left opened for solvent to evaporate before
re-homogenizing. In this way, a sequence of samples, some very close
in composition, could be prepared.

\subsection{Light scattering methods}

The slight difference in the refractive indices of PMMA and {\it
cis}-decalin (1.49 and 1.48 respectively) was enough to render all of
our samples turbid (transmission coefficients $\approx$ 20--40\%). We
therefore used two-color light scattering to extract the
singly-scattered component. The detailed experimental arrangement and
data analysis for this method have been described elsewhere
\cite{two-colour}. Here we just summarize relevant procedures.

Two lasers of different wavelengths, blue ($\lambda_B=488$ nm) and
green ($\lambda_G=514.5$ nm), and two detectors with filters were used
in what were essentially two separate but simultaneous scattering
experiments on the same scattering volume. The incident and scattered
beams were arranged such that the scattering angles $\theta_B$ and
$\theta_G$ were different but the scattering vectors were identical,
i.e. $\vfq_B=\vfq_G=\vfq$, where $\left| \vfq \right|
=4n\pi\sin(\theta/2)/\lambda$ and $n$ is the refractive index of
cis-decalin. The outputs of the two detectors were cross-correlated to
give the intensity correlation function (ICF):
\begin{equation}\label{eq:ICF}
\gc(q,\tau)=\frac{\left<I_B(q,0)I_G(q,\tau)\right>}
{\left<I_B(q)\right> \left<I_G(q)\right>} \;,
\end{equation}

where $I_B$ and $I_G$ are the scattered blue and green intensities
respectively, and the angled brackets denote ensemble averages.

In this arrangement, it can be shown \cite{two-colour} that only the
singly-scattered light of each color probes exactly the same Fourier
component of the density fluctuations and thus are correlated. All
other, multiply-scattered, light does not probe the same component for
both colors and is completely uncorrelated and thus does not
contribute to the time-dependence of $\gc(q,\tau)$. This can be
expressed \cite{two-colour} in terms of the normalized single
scattering dynamic structure factor, $f(q,\tau)$:
\begin{equation}\label{eq:g2}
\gc(q,\tau)= 1 + \beta^2\beta_{MS}^2\left[f(q,\tau)\right]^2\;,
\end{equation}
where the factor $\beta^2$ depends on the ratio of detector area and
coherence area for single scattering and also on the overlap of the
scattering volumes probed by each color. This factor is instrument
related and dependent on scattering angle but not on the sample
used. The other factor $\beta^2_{MS}$ reflects the fraction of
singly-scattered intensities, $\left<I_B^S\right>$ and
$\left<I_G^S\right>$, relative to the total (singly and multiply)
scattered intensities:
\begin{equation}
\beta_{MS}^2 = \frac{\left<I_B^S\right>\left<I_G^S\right>}
{\left<I_B\right>\left<I_G\right>}\;.
\end{equation}
The concentration of polymer in our samples is low. The highest ratio
of intensity scattered from polymer to that from colloid was measured
to be $4\times10^{-3}$. This highest ratio only applied for one sample
(H in Fig.~\ref{fig:phasediag}) at the lowest scattering
angle. Therefore we assume that the scattered intensity is from
colloids only. Under these conditions, our measurements probe the
structure and dynamics of the particles alone.

The static structure factor was measured with the procedure described
in \cite{measureSq}. First, the total average intensities,
$\left<I_B\right>$ and $\left<I_G\right>$, and the intercept,
$\beta^2\beta_{MS}^2=\gc(q,0)-1$, of a concentrated sample of interest
were measured at different scattering vectors $q$. The sample was
rotated continuously during the measurement to ensure ensemble
average. Since the rotation only changes the time-dependence of
$\gc(q,\tau)$, the intercept and average intensity were not
affected. It was found that it was necessary to average measurements
at different height in the sample for non-ergodic samples to reduce
random noise from the finite number of speckles sampled. Then the same
measurements were made on a dilute suspension of known $\phi$ to
obtain the single-particle form factor. The volume fraction of this
dilute sample was $\phi_{\rm dil}=0.01$, small enough that multiple
scattering can be ignored, so that the measured intercept contains
only the instrument related factor: $\beta^2=\gc_{\rm dil}(q,0)-1$,
which was the same as that in the measurement of the concentrated
sample.

The static structure factor is the ratio of singly-scattered intensity
per particle from the concentrated sample to that from the dilute
sample:
$S(q)=(\left<I^S\right>/\phi)/(\left<I_{dil}\right>/\phi_{dil})$. This
was calculated by taking into account multiple scattering and
attenuation of light through the sample:
\begin{equation}\label{eq:Sq}
S(q) = \frac{\phi_{dil}}{\phi}\frac{\sqrt{T_{B,dil} T_{G,dil}}}{\sqrt{T_B T_G}}
\frac{\sqrt{\left<I_B\right> \left<I_G\right>\beta^2\beta_{MS}^2}}
{\sqrt{\left<I_{B,dil}\right>
\left<I_{G,dil}\right>\beta^2}}\;,
\end{equation}
where $T$ is the transmission coefficient of the sample, and subscripts
$B,G$ are for blue and green light respectively.

Dynamic light scattering aims to measure the dynamic structure factor
(DSF), $f(q,\tau)$. This can be extracted from normalizing the ICF
using Eq.~\ref{eq:g2}:
\begin{equation}\label{eq:DSF}
f(q,\tau) = \sqrt{\frac{\gc(q,\tau)-1}{\gc(q,0)-1}}\;.
\end{equation}
However, since most of our samples were either non-ergodic or had very
slow relaxation times, the time-averaged ICF only measured
fluctuations in a small subspace of the whole configuration
space. Explicit ensemble averaging was therefore required, and was
performed by two methods. For short times ($10^{-7}\;\mbox{s} < \tau <
20\; \mbox{s}$), brute-force ensemble averaging was done. Several
hundred (typically between 500-865) of time-averaged ICF's,
$\gc_t(q,\tau)$, and associated scattered intensities, $I_{Bt}$ and
$I_{Gt}$, were measured, each for a duration of 40-60~s. Between each
measurement, the sample was rotated by a small angle to a different
position so that each time-averaged ICF sampled a different
Fourier component. The ensemble-averaged ICF was then constructed as:
\begin{equation}
\gc(q,\tau) = \frac{\left<I_{Bt}I_{Gt} \gc_t(q,\tau)\right>}
{\left<I_{Bt}\right> \left<I_{Gt}\right>}\;.
\end{equation}
For longer times ($\tau > 1$ s), echo DLS was used. Details will be
given elsewhere \cite{echoDLS}. It essentially involves ensemble
averaging by rotating the sample continuously at a constant speed and
correlating the intensities at a small range of delay times around
exact multiples, $\tau \approx nT$, of the rotation period, $T$, where
the correlation function shows `echo' peaks. The rotation decorrelates
the ICF very quickly at small $\tau$. However, after a whole number of
revolutions, the sample comes back to the same orientation and the ICF
recovers to a value that is dependent only on the dynamics of the
particles in the sample over that period of time. This gives rise to
peaks in the ICF at $\tau=nT$. The maxima of these peaks follow the
ensemble-averaged dynamics of the sample because the obtained ICF is
an average over thousands of independent speckles per revolution. We
also correct for imperfections in the rotation using the area under
each echo instead of the echo maximum \cite{echoDLS}. The corrected
ICF at $\tau_n = nT$ was calculated from the measured ICF
$\gc_m(q,\tau)$ as:
\begin{equation}
\gc(q,\tau_n) = \frac{A(\tau_n)}{A(\tau_0)}\left[\gc_m(q,0)-1\right] + 1\;,
\end{equation}
where $A(\tau_n)$ is the area under the echo around $\tau_n=nT$,
$A(\tau_n) = \int(\gc_m(q,\tau)-1)d\tau$. The DSF was then obtained
from corrected ICF in the usual way from Eq.~\ref{eq:DSF}.

We used echo DLS to measure dynamics in the range $\tau=1-10^4$
seconds. Since the rotation used introduces slightly different
alignment in the DLS setup (hence different $\beta^2$), the
resulting intercepts are different from those obtained by brute-force
ensemble averaging. Therefore we scaled the intercept of the echo DLS
results by an arbitrary factor (in the range of 1--2) so that the
resulting DSF from both methods matched in the region of overlap.

\section{Results and discussion}
\subsection{Phase diagram}

The equilibrium phase diagrams of colloid-polymer mixtures at
different values of $\xi$ are well known \cite{cpphaseexpt}. The
non-equilibrium behavior of systems with $\xi \approx 0.1$ at low
volume fractions ($\phi<0.2$) has been studied before
\cite{lowPhiCP1,lowPhiCP2}. Here we concentrate on the higher volume
fraction region ($\phi\ge0.3$).

Many samples were prepared in a range of compositions of
interest. After being homogenized by prolonged tumbling, samples were
left undisturbed for observation. Because the sizes of colloidal
particles are similar to wavelengths of visible light, colloidal
crystals can be seen with the naked eye as iridescent specks.

\begin{figure}
\begin{center}
\epsfig{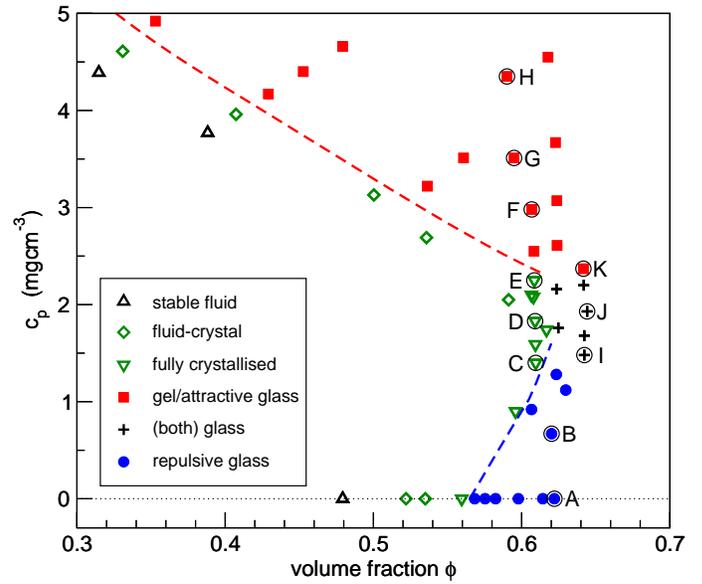}
\caption{Equilibrium and non-equilibrium behaviors of a
colloid-polymer mixture of $\xi=0.09$. Open symbols are those that
reached thermal equilibrium (fluid, fluid-crystal coexistence, and
fully crystallized). Other samples did not crystallize: some showed
characteristics of hard-sphere glasses at the onset of sedimentation
(filled circles), some showed those of attraction-driven glasses and gels
(filled squares), and some showed both (pluses). Dashed curves are
guides to the eye showing the observed boundary where crystallization
ceased. Light scattering data for marked samples labeled A--K are
shown in the following figures. This diagram has been shown in
\cite{unsticking}.}
\label{fig:phasediag}
\end{center}
\end{figure}

Our observations are shown in Fig.~\ref{fig:phasediag}. (These
observations have been presented and briefly discussed before
\cite{unsticking}.) In agreement with equilibrium theory \cite{cptheory}
for systems with short-range attraction, we observed an expansion of
the fluid-crystal coexistence region upon increasing polymer
concentration (diamonds). To the left of this region is a stable fluid
phase (triangles) and to the right is the fully crystalline phase
(inverted triangles). These observations also agree with previous
experiments on similar systems \cite{cpphaseexpt}.

However, samples with very high colloid volume fractions and/or
polymer concentrations (filled circles, squares and crosses) failed to
crystallize for weeks to months even though equilibrium statistical
mechanics predicts either fluid-crystal coexistence or full
crystallinity. Samples with high colloid volume fractions and low
polymer concentrations (circles in Fig.~\ref{fig:phasediag}) showed
all the characteristics of hard-sphere colloidal glass
\cite{HSNature}. Weeks after homogenization and left undisturbed,
sedimentation showed its effect: very thin layers at the top of the
samples developed heterogeneous crystals due to the boundary effect of
the meniscus and gravity. Samples denoted by squares in
Fig.~\ref{fig:phasediag}, with high polymer concentration and moderate
colloid volume fraction, showed signs of transient gels. They collapse
under gravity after some `latency time' as observed previously in
similar systems \cite{lowPhiCP1,lowPhiCP2}. However, the amount of
collapse decreased and transient time increased dramatically in higher
volume fraction samples. For concentrated samples with colloid volume
fraction above 0.55, it took more than 4 weeks to see tiny collapses
of less than half a millimeter at the very top of the meniscus. These
collapses were distinguished from normal sedimentation by their
characteristic sharp and non-flat boundary between the collapsed
material and a clear supernatant. No crystallization was observed in
these samples however long they were left undisturbed. Interestingly,
for non-crystallizing samples with very high colloid volume fraction
and polymer concentration (crosses), characteristics of both
hard-sphere glass and transient gels were present. After 4--8 weeks,
tiny collapses were seen, and also a thin layer of crystal phase
appeared just under the collapsing boundary.

Consider a sequence of samples of similar colloid volume fraction and
increasing attraction, for example samples A-H in Fig.
\ref{fig:phasediag} with $\phi\sim0.6$. According to thermodynamic
equilibrium theory, all these samples should crystallize
\cite{cptheory}. Sample A without polymer was a glass. Sample B with
a small amount of polymer was also a glass as no homogeneous
crystallization was observed for 4 weeks and only heterogeneously
nucleated crystals at the meniscus were observed after 13
days. However, sample C with $\sim 1.4 \;\text{mg cm}^{-1} $ of
polymer completely crystallized in 1 day. This means the glass
transition line has moved to higher $\phi$. Failure to crystallize was
seen again for samples with polymer concentration above $\sim 2.5 \;
\text{mg cm}^{-1}$ (samples F,G,H). The behavior of all the samples in
this region taken together show that the line of failure to
crystallize had a re-entrant shape.

In pure hard-spheres, crystallization ceases at essentially the same
volume fraction as where $f(q,\infty)$ first becomes non-zero, i.e. at
the glass transition \cite{HSNature,HSglass}. If this coincidence
still holds for attractive hard-spheres systems, then we have observed
a re-entrant glass transition in hard spheres with short-range
attraction.

Previous studies of sticky hard spheres by MCT
\cite{glassMCT,gelMCTHiPhi} and computer simulation
\cite{reentrantGlass,reentrantSim} suggest that the re-entrant behavior
is due to two different mechanisms of glassy arrest. The heuristic
picture is as follows. In the `repulsion-dominated' hard-sphere glass,
particles are caged by their neighbors at high enough volume
fraction. Short-range attraction clusters the particles of the cage
and opens up holes, ultimately melting the glass. However, increasing
the attraction further leads to an `attraction-dominated' glass where
particles stick to their neighbors with long-lived bonds. In this
terminology, samples A and B are repulsive glasses and F-H are
attractive glasses. Samples I--K must lie in the region where these
two types of glass merge as they show characteristics of both types, with
further evidence in the dynamics shown in section \ref{sec:DSF}.  The
next sections, with results from light scattering, will give insights
into the structure and dynamics of these glasses, and the nature of
the re-entrant transition between them.

\subsection{Static structure factor}

\begin{figure}
\begin{center}
\epsfig{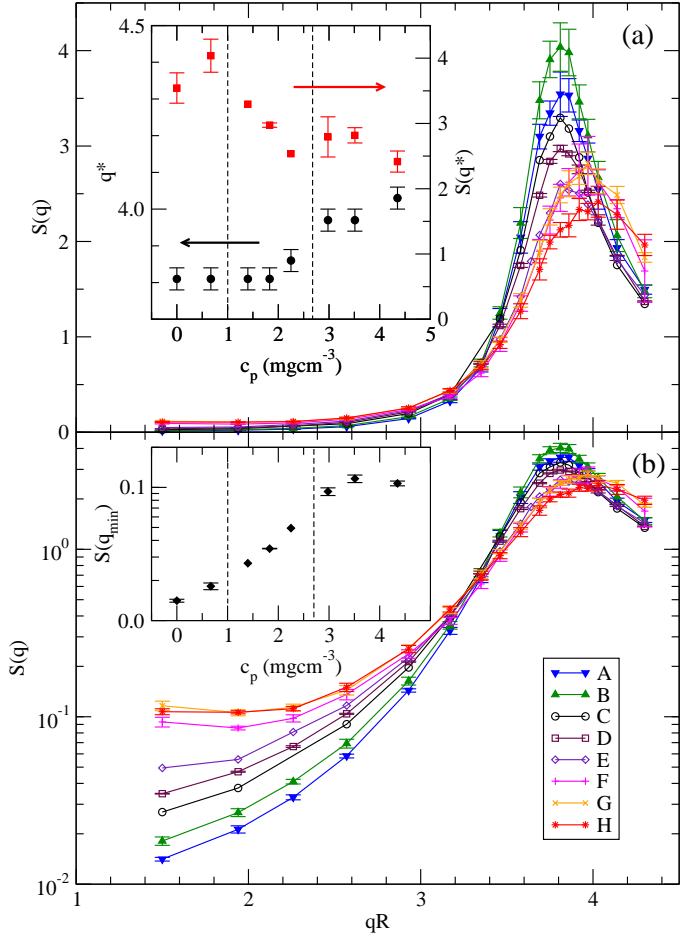}
\caption{Static structures factor of samples A--H ($\phi\sim0.6$) as
function of dimensionless wave vector $qR$. The lines are guides to
the eye. (a) The peak position $q^*$ shifts to higher $q$, while its
height reduces and width increases upon increasing attraction. The
inset shows the peak positions and heights of these static structure
factors as a function of polymer concentration. (b) The same SSFs
plotted with logarithmic vertical axis shows the increase of $S(q)$ at
low $q$. The inset shows $S(q)$ at the lowest wave vector
$qR=1.50$. Vertical dashed lines in both insets indicate the glass
transitions observed in Fig.~\ref{fig:phasediag}.}
\label{fig:sqLo}
\end{center}
\end{figure}

We measured the static SSFs of the samples whose symbols are circled
in Fig.~\ref{fig:phasediag}. Note that samples C--E were measured as
metastable fluids, i.e. before any crystal nucleation took place. 
Consider first the results for a sequence of samples (A--H) with 
$\phi\approx 0.6$, Fig.~\ref{fig:sqLo}. These samples span the
re-entrant glass transition line where the crystallization behavior
showed dramatic changes. However, no re-entrant behavior can be seen
in the SSF. Instead, there are only gradual changes upon increasing
the attractive interaction. These gradual changes have been predicted
by theory \cite{glassMCT}, and observed before in other experimental
systems \cite{glassMicellar3}.

The most obvious and most easily quantifiable changes are in the
height and position of the main peak. Broadly speaking, and taking
experimental uncertainties into account, the peak reduces in height
and shifts to higher $q$ when the attraction is increased (inset
Fig.~\ref{fig:sqLo}(a)). In detail, the peak position, $q^*$,
remains constant (at $q^* R \approx 3.8$, samples A--D) until just
before we enter the attractive glass region (sample E), whereupon it
increases by $\approx 5\%$ to reach another constant value ($q^* R
\approx 4$, samples F--H). These samples have approximately constant
$\phi$ (in fact it decreases slightly from A to H,
Fig.~\ref{fig:phasediag}). The increase in $q^*$ is the result of a
significant fraction of neighboring particles becoming trapped in each
others' narrow depletion potential well when the attractive glass
forms. Quantitatively, a 5\% increase in $q^*$ corresponds to a 15\%
increase in the local packing fraction, from 0.6 to 0.69; the latter
is the random close packing volume fraction for our system (measured
by spinning down a sample of known $\phi$). In other words, the nearest
particles in the attractive glass are practically touching.

The clustering of particles at constant volume necessarily implies
that the average number of nearest neighbors should decrease, and that
`holes' are opened up to render the structure more inhomogeneous on
the spatial scale of a few particles. The former is reflected in the
decrease in $S(q^*)$. Significantly, upon increasing the attraction
from zero, the decrease in the peak height starts at the point of the
melting of the repulsive glass, and continues until we enter the
attractive glass region, whereupon the peak height remains constant
(inset Fig.~\ref{fig:sqLo}(a)).

The increased heterogeneity is reflected in a rise in the SSF at low
$q$, Fig.~\ref{fig:sqLo}(b). The smallest $q$ we have studied was
$q_{\text{min}}R=1.50$, corresponding to a length scale of about 4
particle radii. The value of $S(q_{\text{min}})$ increases nearly
exponentially with the polymer concentration between samples A--E
(inset Fig.~\ref{fig:sqLo}(b)), and thereafter remains constant. The
increased density fluctuations at this length scale corresponds to the
opening up of `holes' due to particle clustering.

Note that all three features considered, $q^*$, $S(q^*)$ and
$S(q_{\text{min}})$, remain virtually constant for all three
attractive glass samples, F--H. Once particles drop into each others'
narrow attractive potential wells, any further {\it structural}
changes will be hard to resolve. We shall see, however, that the {\it
dynamics} continues measurably to evolve from sample F to sample H: in
this regime of almost-touching nearest neighbors, a very small change
in the structure has very large dynamic consequences.

\begin{figure}
\begin{center}
\epsfig{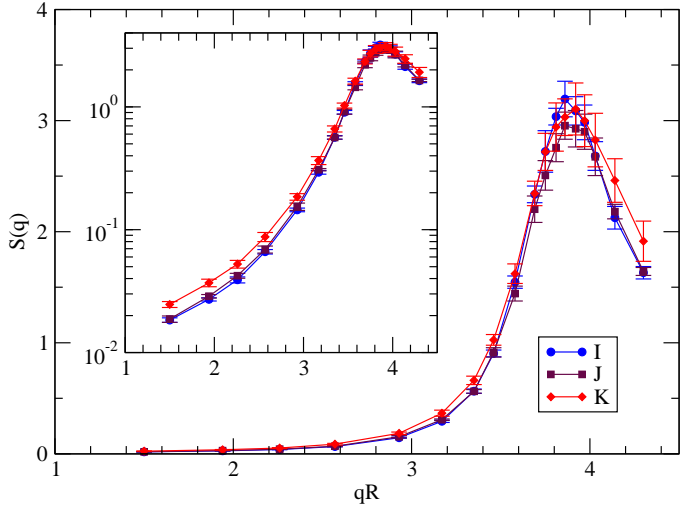}
\caption{Static structure factors of samples I--K with
$\phi \sim 0.64$. The inset shows the same data with a logarithmic
vertical axis.}
\label{fig:sqHi}
\end{center}
\end{figure}

All the qualitative features we observed in the evolution of the SSFs
for samples A--H are also seen in the SSFs for samples I--K at the
higher volume fraction of $\phi\approx 0.64$,
Fig.~\ref{fig:sqHi}. However, the effects are significantly less
obvious, largely because the range of polymer concentration is now
much smaller and $\phi$ is higher. At low $q$, the values of
$S(q_{\text{min}})$ are lower than those of similar polymer
concentration but lower $\phi$ (C--E) (Fig.~\ref{fig:sqLo}(b)). This
is because at higher volume fraction, a tight local clustering of some
particles does not create so much room elsewhere --- there is less
space for developing heterogeneities.

\subsection{Dynamic structure factor} \label{sec:DSF}

\begin{figure}
\begin{center}
\epsfig{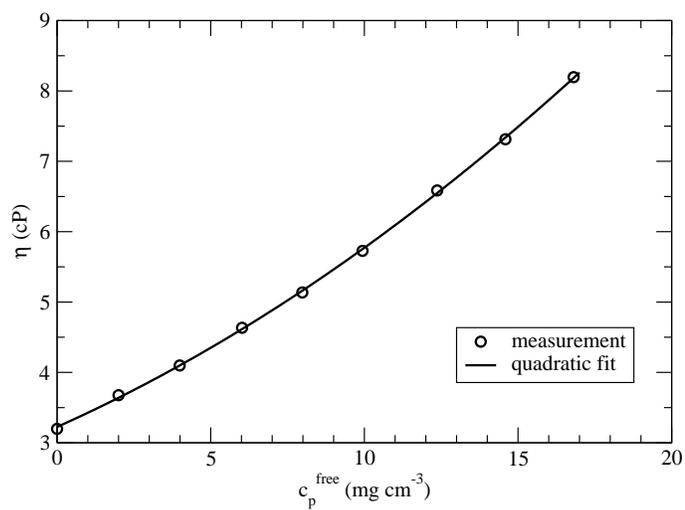}
\caption{The viscosity of polystyrene in {\it cis}-decalin at $20^\circ$C
at different polymer concentrations. A quadratic fit (solid line) was
used to interpolate to viscosities of samples with different \cpfree.}
\label{fig:viscosity}
\end{center}
\end{figure}

Our goal is to study how the polymer-induced depletion attraction
affects the particle dynamics. But the presence of the polymer
influences the dynamics in another, essentially trivial, manner --- by
increasing the effective viscosity of the medium in which the
particles diffuse from that of the pure solvent, $\eta_0$, to that of
a polymer solution, $\eta_r \eta_0$ at concentration \cpfree. To
determine $\eta_r$, we measured the viscosity of pure polymer
solutions with a miniature suspended-level viscometer,
Fig.~\ref{fig:viscosity}, and used a quadratic fit to the data to
obtain $\eta_r$ for our samples. The value of \cpfree in each sample
was estimated from $c_p$, $\phi$ and $\xi$ using an approximate
expression based on scaled-particle theory \cite{cptheory}.

The rate of dynamical decay at wave vector $q$ depends on the length
scale being probed; in dilute systems it scales as $q^2$. Thus, in
order to compare the dynamics of different samples at different wave
vectors, and to highlight the effects of the attraction, we scaled the
delay time variable of the DSFs by the relative viscosity $\eta_r$ and
the dimensionless wave vector $(qR)^2$, so that DSF is presented as a
function of the `scaled time' $(qR)^2\tau/\eta_r$. Note that for the
lowest $q$ studied, the scaled time is very close to the real time,
while at the highest $q$, it is increased by about an order of
magnitude.

\begin{figure}
\begin{center}
\epsfig{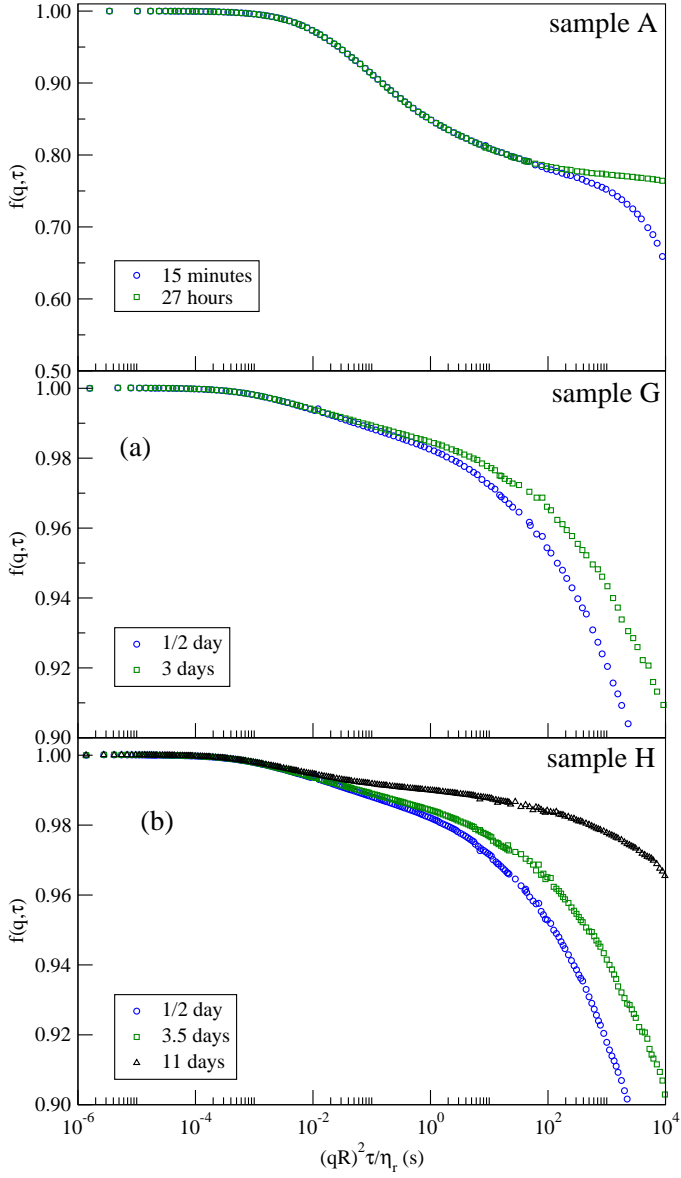}
\caption{Aging in samples A, G and H at $qR=2.93$. The legend indicates
the waiting time between the end of tumbling and the start of
measurements. The DSFs slow down and the points of inflection become
clearer with increasing age of the samples.}
\label{fig:aging}
\end{center}
\end{figure}

We found aging \cite{generalAging} in all non-crystallizing
samples. The dynamics slowed down with the `waiting time' --- the time
interval between the cessation of tumbling and the beginning of
measurements, Fig.~\ref{fig:aging}. It is known that the hard-sphere
glass ages \cite{HSaging}. We found that the rate of aging in
different glasses were different and that its effects were
complex. Repulsive glasses aged only in the first day or two, after
which they did not evolve within the time window of the
measurements. Attractive glasses, on the other hand, showed different
dynamics with age for up to 10 days. Aging is complicated enough to be
the subject of a separate study and was not investigated
systematically in this work. To eliminate as much as possible aging
effects on dynamical results within practical limits of waiting time,
we present DSFs of glassy samples with age between 1 and 4 days. The
dynamics of crystallizing samples (C--E) were measured while they were
in the metastable state well before the appearance of
crystallization. Below we first show results of different samples at
the same $q$, then at different $q$ for the same sample.

\subsubsection{Constant scattering vectors, variable compositions}

\begin{figure}
\begin{center}
\epsfig{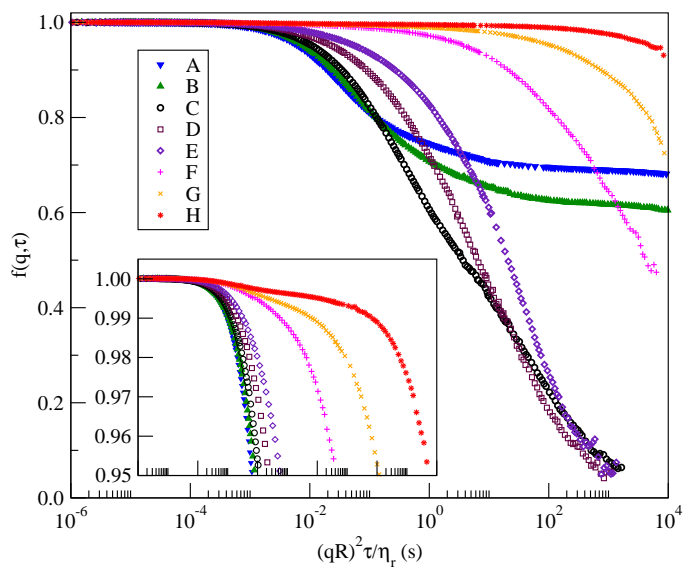}
\caption{Collective dynamic structure factors at $qR=1.50$ from
samples A--H spanning the re-entrant region. The time axis is
scaled to dimensionless length scale $(qR)^2$ and relative polymer
solution viscosity $\eta_r$. The inset shows the same plots on an
expanded vertical axis.}
\label{fig:up0.6q23}
\end{center}
\end{figure}

The DSFs of samples A--H at $qR=1.50$, Fig.~\ref{fig:up0.6q23},
clearly evolve non-monotonically with increasing polymer concentration
and show re-entrant behavior. Briefly, samples A and B are non-ergodic
within our time window, while samples C--E are ergodic (their DSFs
decay completely to zero), and samples F--G become non-ergodic again.

In detail, the DSF of sample A, a pure hard-sphere glass, shows a
plateau at $f_A(q,\infty)\approx 0.7$, corresponding to particles
getting `stuck' in their nearest-neighbor cages. This can be compared
with previous work \cite{HSglass,HSnonergodicity}. Note that in doing
so, it is important to compare samples with the same density {\it
relative to random close packing}: i.e. the same $(\phi_{\rm
rcp}-\phi)/\phi_{\rm rcp}$, since $\phi_{\rm rcp}$ differs according
to the polydispersity of the colloids \cite{polyrcp}.

With a small amount of polymer added to the hard-sphere glass, sample
B shows the same qualitative dynamics. Quantitatively, however, the
height of the plateau is lower, $f(q,\infty) = 0.62$. This indicates
that particles in B are not as restricted as in A, i.e. the cage is
loosened by the attractive interaction, but still remains closed in
our time window.

The DSF of sample C decayed completely in (a `scaled time' of) about
1000 seconds, as did those for the other crystallizing samples D and
E. It is interesting to note that the DSFs of these three samples slow
down upon increasing polymer concentration but all reach zero at about
the same scaled time. The DSF of sample C shows the remnant of a plateau at a
scaled time of $ \approx$~10~s. The DSFs for samples D and E exhibit a
very stretched {\it single} decay, rather than a two-stepped
process. This is unusual behavior for a fluid at volume fraction $\phi
\sim 0.6$ (at least at first sight).

The intermediate, $\beta$, and long-time, $\alpha$, decay in a dense
hard-sphere fluid are attributed to particles `rattling' in their
local neighbor cages, and escaping from these cages, respectively
\cite{HSglass}. Attraction hinders the `rattling' by trapping
particles in potential wells, but accelerates the cage opening by
clustering. At some polymer concentration (or attraction strength),
the two time scales coincide. If at this point the attraction alone is
not enough to trap the system in an non-ergodic state, we will observe
the melting of the repulsive glass into an ergodic fluid dominated by
attraction. This is the case for sample C, where the $\alpha$ and
$\beta$ decays are barely distinguishable in the DSF. At higher
polymer concentrations, the cage concept is no longer appropriate for
describing the particle dynamics --- for it to be valid, a particle
has to `rattle' many times in a cage before it opens. Note that this
is a distinctive feature of dense fluids with {\it short-range}
attraction. In a dense fluid with {\it long-range} attraction, the
effective potential well experienced by any particle due to its
neighbors is essentially flat. This adds a (negative) constant to the
free energy, so the phase behavior \cite{widom} and dynamics of the
system are still controlled by repulsion (or, equivalently, entropy).

Note that the shape of the DSF of sample C at $qR = 1.50$ is similar
to that shown in curve~2, Fig.~11 of \cite{glassMCT}. This DSF was
calculated at $qR = 2.1$ for a sample in the re-entrant portion of the
state diagram in a system that {\it just} shows a glass-glass
transition and an A3 point. Recent calculations for colloid-polymer
mixtures \cite{gelMCTHiPhi} suggests that our system, with $\xi
\approx 0.09$, should show exactly these features.

The DSFs of samples F--H are, once more, non-ergodic in our time
window: they do not decay completely even after $10^4$ seconds. Simple
extrapolation indicates that it would take these DSFs at least $10^6$
seconds to reach zero. The DSFs of samples G and H show points of
inflection; that for sample H is clearer and occurs at $f = 0.995$ ---
a very high value compared to the plateaus in hard-sphere
glasses. These high points of inflection can be associated with
dynamics originating from particles rattling in very narrow attractive
potential wells.

\begin{figure}
\begin{center}
\epsfig{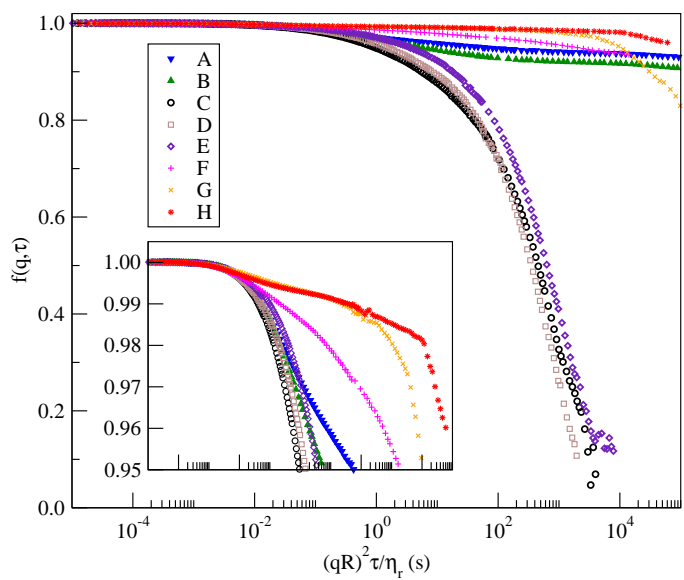}
\caption{DSFs at the peak of the SSFs for samples A--H. The inset shows
the same plots on an expanded vertical axis.}
\label{fig:up0.6qp}
\end{center}
\end{figure}

\begin{figure}
\begin{center}
\epsfig{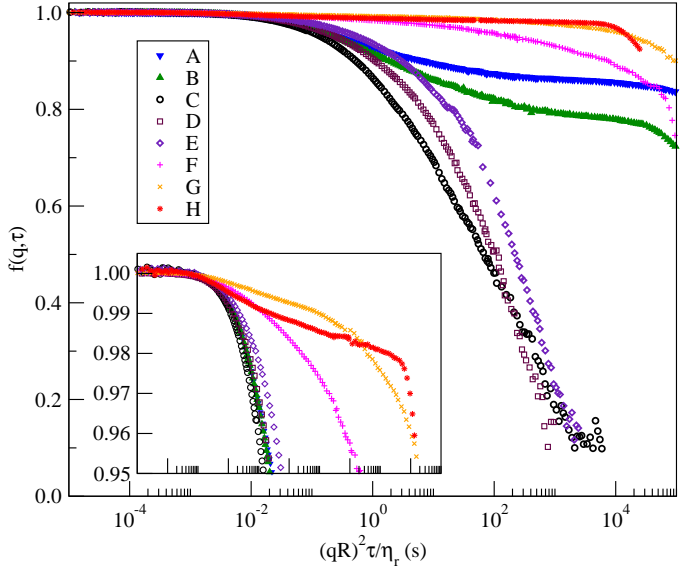}
\caption{DSFs of samples A-H at $qR=4.30$, to the right of all $S(q)$
peaks. The inset shows the same plots on an expanded vertical axis.}
\label{fig:up0.6q70}
\end{center}
\end{figure}

At other wave vectors, Figs.~\ref{fig:up0.6qp} and
\ref{fig:up0.6q70}, the DSFs behave in a similar way, namely
relatively low plateaus in the repulsive glasses A and B, complete
decay in the metastable fluids C--E, and extremely slow dynamics in
the attractive glasses F--H, with very high points of inflection in G
and H. Note, however, that at the peak of the corresponding SSFs, the
DSFs for samples C--E are barely distinguishable (Fig.
\ref{fig:up0.6qp}).

\begin{figure}
\begin{center}
\epsfig{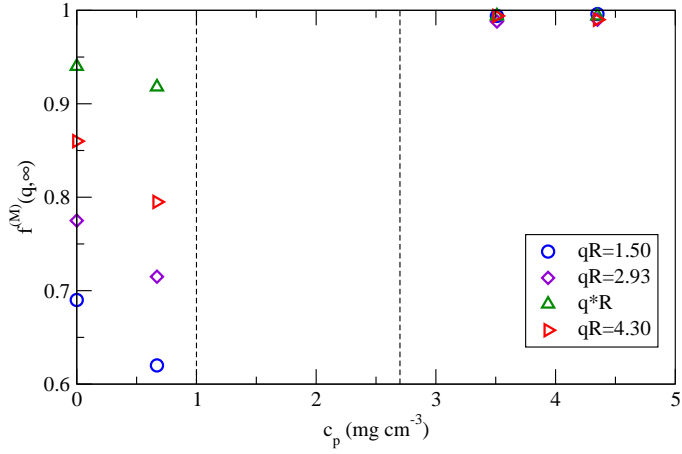}
\caption{The measured non-ergodicity parameters, $f^{(M)}(q,\infty)$,
at different wave vectors as a function of polymer concentration in
samples (left to right) A, B, G, and H. The dashed lines indicate the
glass transitions observed in Fig.~\ref{fig:phasediag}.}
\label{fig:inftycp}
\end{center}
\end{figure}

The plateaus in the DSFs of the repulsive glasses can be used as a
measure of $f(q,\infty)$, the non-ergodicity parameter. An estimate of
this quantity for the attractive glasses is more problematic, partly
due to significant aging in our time window. To proceed, we use the
value of $f$ at the point of inflection as a surrogate; we call this
the `measured' $f(q,\infty) \equiv f^{(M)}(q,\infty)$. The evolution
of $f^{(M)}(q,\infty)$ with increasing polymer concentration (samples
A--H) is shown in Fig.~\ref{fig:inftycp}. The non-ergodicity
parameter decreases slightly when moving from A to B, away from the
hard-sphere glass. When attraction melts the repulsive glass,
$f^{(M)}(q,\infty) = 0$ for samples C--E (not shown). Sample F did not
crystallize and showed non-ergodic dynamics up to $10^4$ seconds but
did not exhibit any discernible point of inflection in its
dynamics. Samples G and H had extremely high non-ergodicity parameters
of nearly 1. A `jump' in $f(q,\infty)$ when moving from repulsive to
attractive glass was predicted by MCT (Fig.~7 in \cite{glassMCT}).

\begin{figure}
\begin{center}
\epsfig{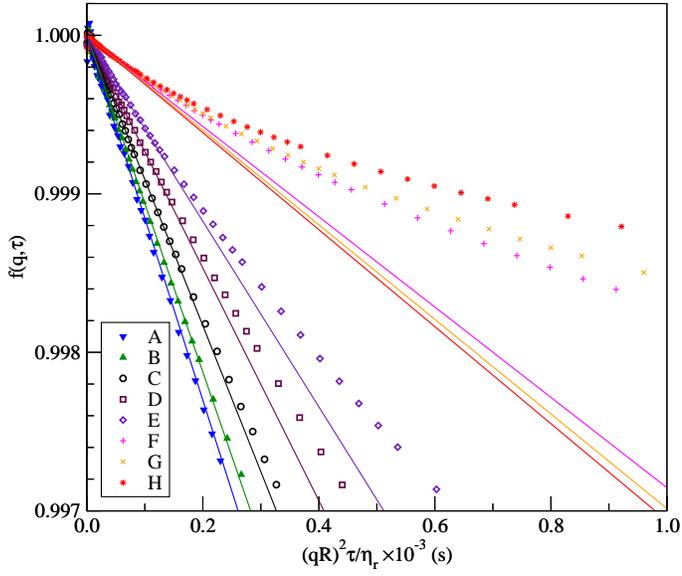}
\caption{The short-time dynamics of samples A--H at $qR=1.50$. The
straight lines are fits to the linear part of the DSFs at
$\tau\rightarrow 0$. The dynamics departs from an initial diffusive regime
progressively earlier upon increasing attraction. The
short-time diffusion coefficient in the limit $\tau\rightarrow 0$ is
also reduced significantly.}
\label{fig:shorttime}
\end{center}
\end{figure}

\begin{figure}
\begin{center}
\epsfig{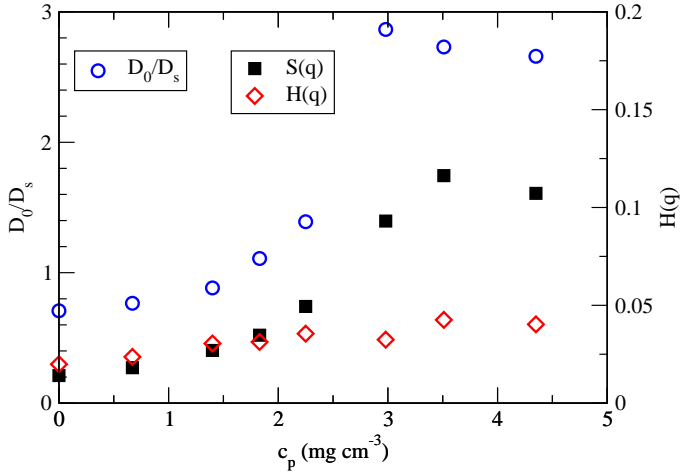}
\caption{The normalized short-time diffusion coefficient $D_0/D_s$(circles,
left scale), static structure factor $S(q)$ and hydrodynamic factor
$H(q)$ (right scale) at $qR=1.50$. $D_s$ and $S(q)$ were extracted
from Fig.~\ref{fig:shorttime} and \ref{fig:sqLo} respectively. The
decrease in $D_s$ is nearly in line with the increase in $S(q)$ so
that $H(q)$ only increased slightly.}
\label{fig:DsHq}
\end{center}
\end{figure}

The evolution of the short-time dynamics of the whole sequence of
samples is also interesting. Fig.~\ref{fig:shorttime} shows the
short-time behavior of the DSFs for A--H at large length scale,
$qR=1.50$, where experimental noise is lowest. Note the very small
vertical interval, 1.000 to 0.997, spanned in this figure; thus only
the first 0.3{\%} of the decays of the DSFs are being analyzed.  The
DSFs of repulsive glasses A and B possessed relatively long linear
parts, corresponding to the first term in $\tau$ in the expression
derived from the Smoluchowski (many-particle diffusion) equation
\cite{puseyreview}: $f(q,\tau)=1-\frac{D_0H(q)}{\eta_rS(q)}q^2\tau +
O(\tau^2)$, where $D_0$ is the free-particle diffusion constant in
pure solvent (with no polymer) $D_0=k_BT/6\pi\eta_0R$, and $H(q)$ is
the hydrodynamic factor. This linear regime of the DSFs indicates that
at short time, individual particles still diffuse freely without the
influence from direct interaction with their neighbors. The change in
limiting slope as $\tau\rightarrow 0$, or the short-time diffusion
coefficient $D_s(q) = D_0 H(q)/S(q)$, can be almost entirely explained
by the change in $S(q)$ (Fig.~\ref{fig:DsHq}), including the strong
decrease on entering the attractive glass regime. What is more
interesting is that the dynamics depart from free diffusion
progressively earlier upon increasing attraction
(Fig.~\ref{fig:shorttime}). In fact, for the attractive glasses
F--H, the particles are confined so tightly by the attractive
potential wells that the DSFs display non-linearity almost immediately
(cf. also insets to Figs.~\ref{fig:up0.6q23}--\ref{fig:up0.6q70}).

\begin{figure}
\begin{center}
\epsfig{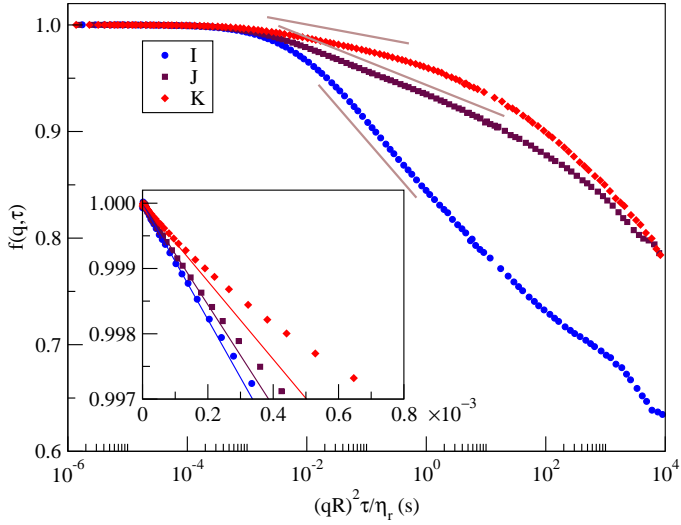}
\caption{The DSFs at $qR=1.50$ for samples I--K with
$\phi\sim0.64$. Extremely stretched relaxation is found in all three
samples with logarithmic decay over long ranges of $\tau$ (straight
lines). The inset shows the short-time dynamics, which deviate from
the diffusive regime from very early times.}
\label{fig:up0.64q23}
\end{center}
\end{figure}

Moving to the (shorter) sequence of samples at the higher volume
fraction of $\phi \approx 0.64$ and closer to the intersection of the
two glass transition lines, samples I--K in
Fig.~\ref{fig:phasediag}, we see the emergence of remarkably
stretched-out, extremely slow dynamics. Consider first the data at $qR
= 1.50$, Fig.~\ref{fig:up0.64q23}. In terms of short-time dynamics
(inset, Fig.~\ref{fig:up0.64q23}), samples I and J are comparable to
samples C and D, while sample K shows a behavior intermediate between
those of samples E and F. At intermediate times, the decay is
linear with respect to the logarithm of the scaled time. Thereafter
there is an incipient plateau at $f \sim 0.7$ in sample I, reminiscent
of the plateau in repulsive glasses A and B, before a further decay,
but never beyond $\sim 0.62$ in our time window. There is no incipient
plateau for the other two samples. Note that the DSF of sample I shows
aspects of the behavior of repulsion-dominated glasses (long time) and
a fluid dominated by short-range attraction (short time). The two
regimes are `bridged' by a stretched log-time decay.

\begin{figure}
\begin{center}
\epsfig{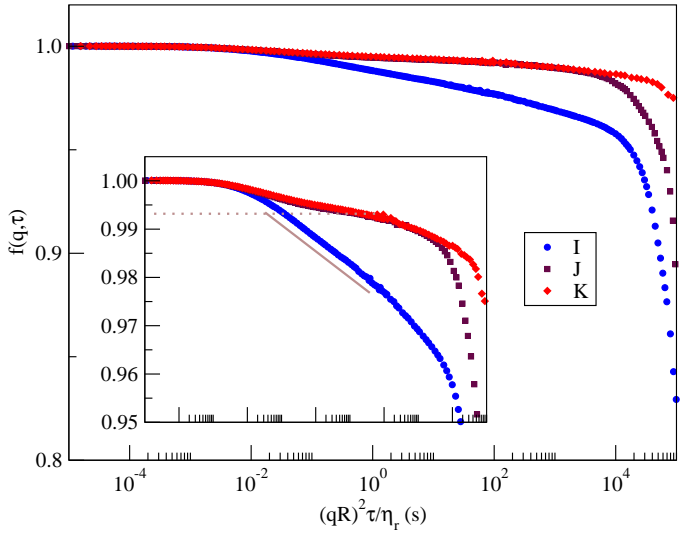}
\caption{DSFs at the peak of the SSH for samples I--K. All
decay much slower than at low $q$. Sample I shows a logarithmic decay
for about 3 decades in scaled time. Samples J and K develop very high
plateaus (inset).}
\label{fig:up0.64qp}
\end{center}
\end{figure}

At the peak of the SSF, Fig.~\ref{fig:up0.64qp}, sample I behaves in
a similar way at short to intermediate times, while there is no
incipient plateau at long times. Samples J and K develop an incipient
plateau as high as $\sim0.993$ (inset Fig.~\ref{fig:up0.64qp})
before turning over to decay more rapidly in logarithmic time.

The fact that these samples show extremely stretched out dynamics,
logarithmic in time, suggests that they are very close to the A3
critical point predicted by MCT, where the repulsive and attractive
glasses become indistinguishable
\cite{glassGlassLogDecay,glassMCT,SciortinoLogtime}. In particular,
the shape of the DSF of sample I at $qR = 1.5$ is comparable to
curve~3 in Fig.~11 of \cite{glassMCT}, calculated at $qR = 2.1$ for
a sample on the repulsive glass transition line very close to where it
intersects the attractive glass transition line for a system that just
shows an A3 singularity. This is not inconsistent with the position of
sample I on the state diagram, Fig.~\ref{fig:phasediag}, of our
system at $\xi \approx 0.09$ \cite{gelMCTHiPhi}.

Heuristically, we may begin to make sense of log-time decays as
follows. At high enough volume fraction, the average distance between
neighboring particles will decrease to a value such that they are
always well within the attraction range of each other \footnote{In our
system the estimated distance between particles for samples I--K
($\phi=0.64$) from random close pack ($\phi_{rcp}=0.69$) is
$(\phi_{rcp}/\phi)^{1/3} = 1.03 \sim 1 + \xi/3$, where the attractive
potential is half of the maximum depth.}.  If the attraction is strong
enough, the restriction of particle movement due to the neighbor cage
and the restriction caused by bonding between particles take place
simultaneously at all times. This competition between two opposite
mechanisms may lead to a broad distribution of decay times and
therefore a very stretched out DSF
\footnote{Formally, a $\tau^{-1}$ distribution of decay times gives a
decay linear in $\log
\tau$. Limitations in our data mean, however, that we cannot use them
to back out the actual decay-time distribution in our samples.}.

\subsubsection{Constant compositions, variable scattering
vectors}\label{sec:dynsameq}

\begin{figure}
\begin{center}
\epsfig{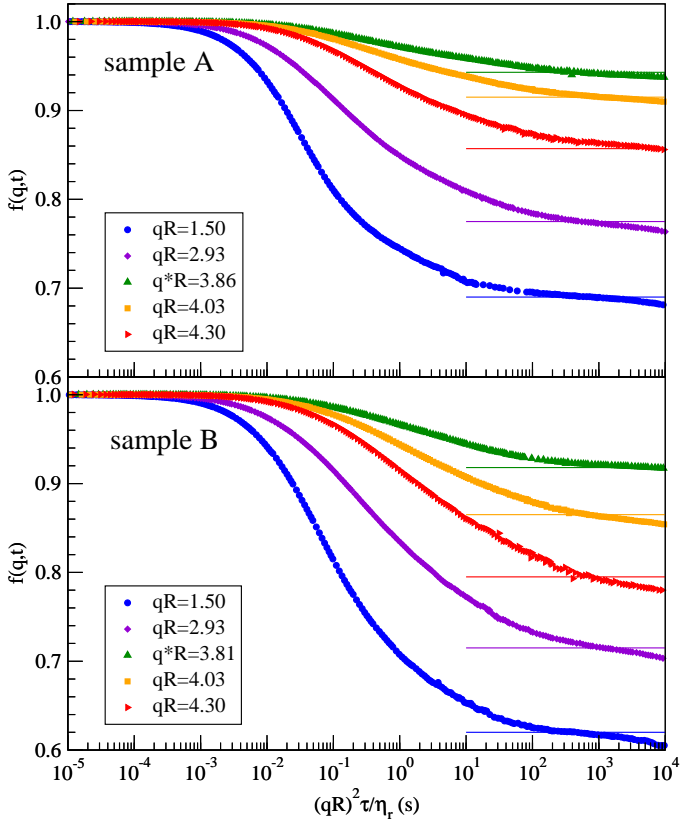}
\caption{The DSFs of samples A and B at different $q$ vectors, $q^*$
denotes the peak position of the static structure factor $S(q)$.  The
general shape of the DSFs are very similar. Horizontal lines denotes
the height of the plauteau (non-ergodicity parameters) that are
plotted in Figs.~\ref{fig:inftycp} and \ref{fig:inftyq}.}
\label{fig:ABqall}
\end{center}
\end{figure}

\begin{figure}
\begin{center}
\epsfig{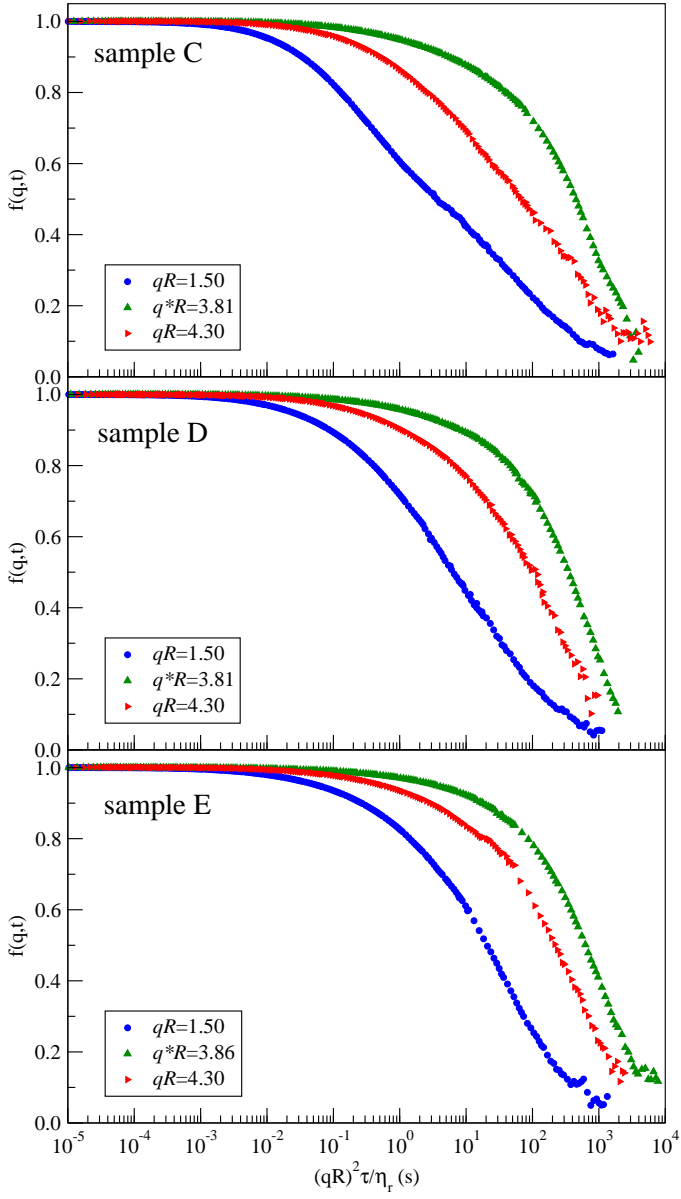}
\caption{The DSFs of samples C--E at different $q$. The rate of
decay varies in the opposite direction to $S(q)$ (c.f. Fig.
\ref{fig:sqLo}). However, all decay to zero at approximately the same
scaled time. Except for sample C at the lowest $q$, all other DSFs do not
show two distinct relaxation processes as other dense fluids.}
\label{fig:CDEqall}
\end{center}
\end{figure}

\begin{figure}
\begin{center}
\epsfig{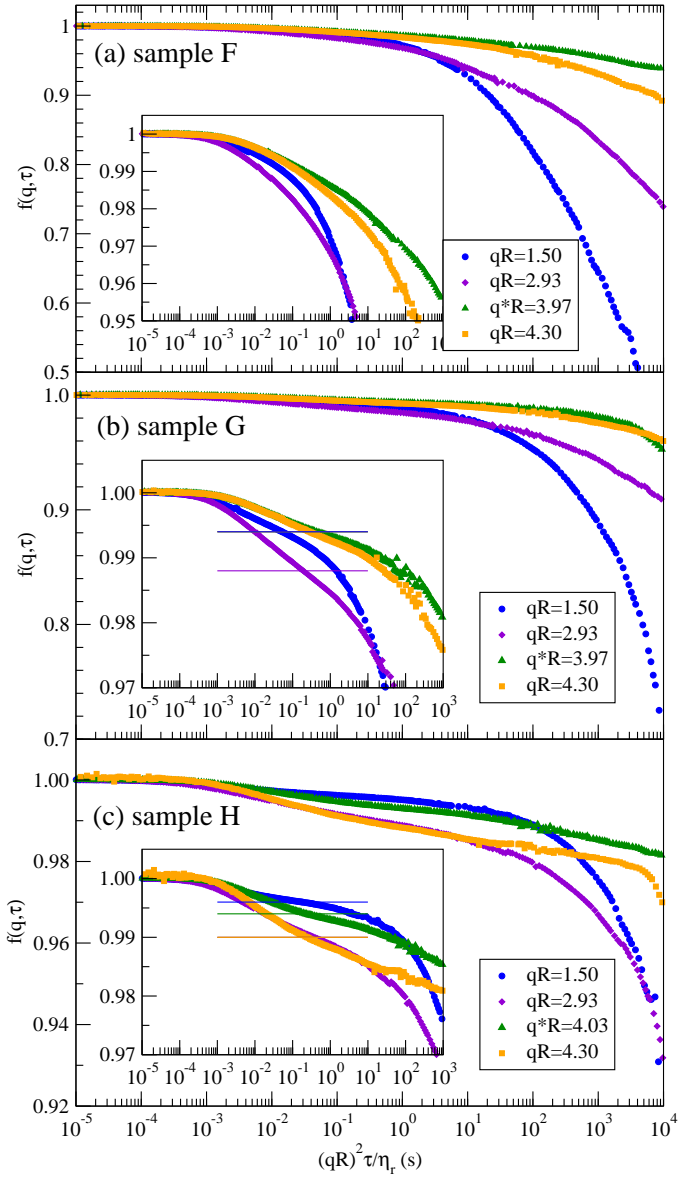}
\caption{The DSFs of samples F--H at different $q$. The vertical
axes span different ranges. Sample F did not show a point of
inflection, but G and H have very high points of inflection
(horizontal lines), the values of which are used in Figs.~
\ref{fig:inftycp} and \ref{fig:inftyq}.}
\label{fig:FGHqall}
\end{center}
\end{figure}

\begin{figure}
\begin{center}
\epsfig{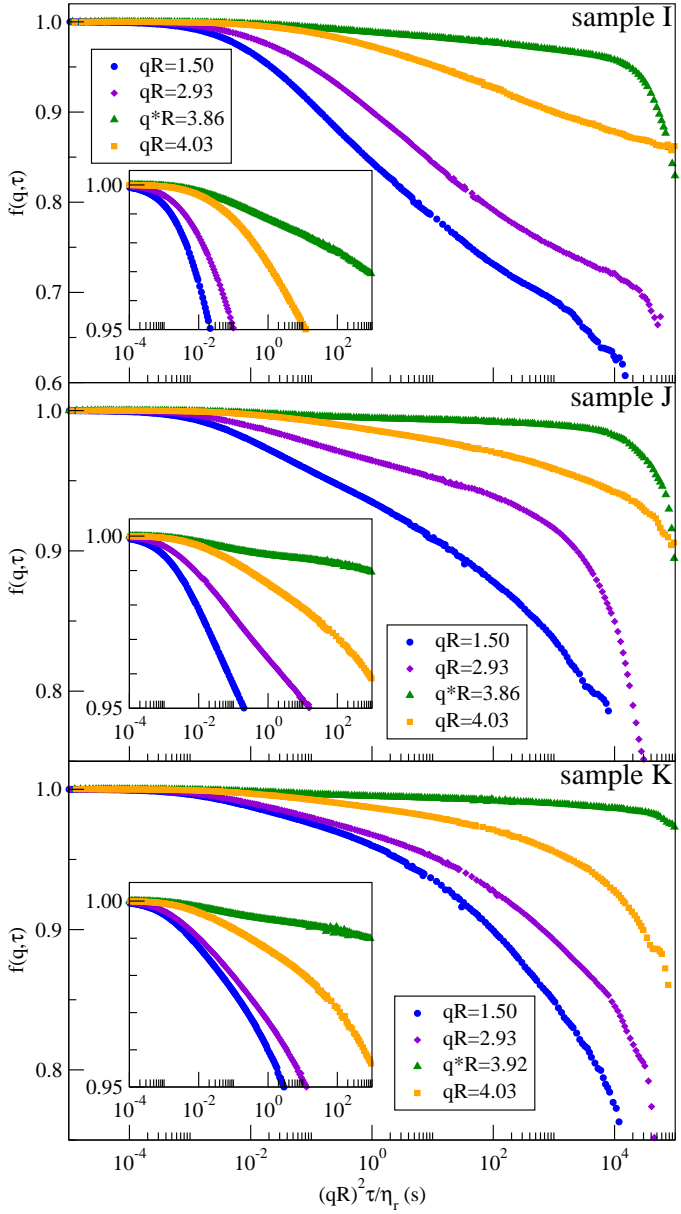}
\caption{The DSFs of samples I, J and K at different $q$. The insets
show the same quantities with expanded vertical axes. The relaxations
show similar behavior at all wave vectors except at the peak of
$S(q)$. Sample I decays to a logarithmic section and then appears to
turn up to a plateau. Sample J shows a very long section of
logarithmic decay. Sample K is similar to J with a shorter stretch of
logarithmic decay. In the early decay at the peak of $S(q)$, the DSF
of sample I has a long stretch of logarithmic decay whereas samples J
and K develop very high plateaus.}
\label{fig:IJKqall}
\end{center}
\end{figure}

In this section, we show for completeness the dynamics of each sample
at different scattering vectors in Figs.~
\ref{fig:ABqall}--\ref{fig:IJKqall}. The change of DSFs with $q$ in
repulsive glasses A and B are in agreement with previous work
\cite{HSnonergodicity,HSglass}. Other samples show the general trend
that the dynamics become slower at scattering vectors with higher
$S(q)$. The only exception concerns the intermediate-time dynamics of
the attractive glasses F--H (insets, Fig.~\ref{fig:FGHqall}). The
significance of the rather complicated $q$-dependence of the
intermediate-time dynamics of these samples is not clear. Nor do we
know of any detailed calculations to date that can throw light on this
issue.

\begin{figure}
\begin{center}
\epsfig{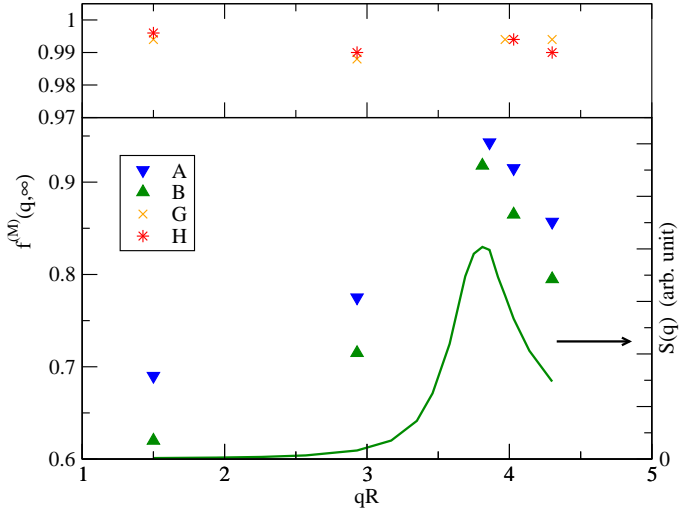}
\caption{The measured non-ergodicity parameters of samples A, B, G and H as a
function of scattering vector $q$ (points), and the static structure
factor of sample B (line) for comparison. The non-ergodicity parameters of
repulsive glasses A and B follow the static structure factor,
whereas those of the attractive glass are extremely high and hardly
fluctuate with $q$ (upper pane with expanded vertical axis).}
\label{fig:inftyq}
\end{center}
\end{figure}

The systematic $q$-dependent data shown in Figs.
\ref{fig:ABqall}--\ref{fig:IJKqall} allow us to investigate the
$q$-dependence of the measured non-ergodicity parameter,
$f^{(M)}(q,\infty)$, in detail. The measured non-ergodicity parameters
of glassy samples A, B, G and H are shown as a function of scattering
vector $q$ in Fig.~\ref{fig:inftyq}. The data for repulsive glasses A
and B vary essentially with the static structure factor, as observed
in hard-sphere glasses \cite{HSnonergodicity}.  Attractive glass G and
H on the other hand showed extremely high measured non-ergodicity
parameters that hardly vary with $q$. This agrees with predictions by
MCT (c.f. Fig.~8 in \cite{glassMCT}).

\section{Conclusion}

We have studied a dense system of hard-sphere colloids with a
short-range inter-particle attraction induced by the depletion effect
of added non-adsorbing polymer. The observed crystallization behavior
as well as particle dynamics studied by DLS reveal a re-entrant glass
transition. With little attraction, the system at high enough volume
fraction is `stuck' in a repulsive glassy state where the arrest is
due to caging by neighboring particles. Our data support the
suggestion \cite{gelGlassYukawa} that attraction causes particles to
cluster, thus opening up holes in the cages and melting the glass. At
the same time, the attraction slows down the particle dynamics. We
found that the repulsive glass melts when the characteristic time of
the attraction-dominated particle dynamics becomes comparable to that
of cage opening. The resulting ergodic fluid shows a distinctive
dynamical feature: despite the fluid's high density, its DSF does not
show distinct $\alpha$ and $\beta$ relaxation processes. Increasing
the attraction further leads to different kind of arrest where the
strong attraction between particles create long-lived bonds and
prevent structural rearrangement, giving rise to an
attraction-dominated glass. Detailed light scattering has been used to
probe the effect of attraction on both structure and dynamics.

Qualitatively, this scenario agrees well with predictions from MCT
calculations (with those reported in \cite{gelMCTHiPhi} being closest
to the present experimental system). In particular, we observed very
slow, log-time dynamics in the DSFs in the region where the two glass
transition lines are expected to meet. Quantitatively, however, our
results stand as a challenge to MCT (or any other theory): the
detailed calculations needed for direct quantitative comparison have
not, to our knowledge, been performed.

A detailed comparison between experiment and theory faces a number of
non-trivial problems. First and foremost, since calculated and
measured glass transition thresholds differ, choices exist as to what
constitute `corresponding state points' for the purpose of making the
comparison. In the case of pure hard spheres, where $\phi_g^{\rm MCT}
\approx 0.52$ and $\phi_{g}^{\rm expt} \approx 0.58$, it is accepted
practice to compare measurements and calculations at the same relative
volume fraction $(\phi - \phi_g)/\phi_g$ \cite{HSglass}. The situation
is more complex in a colloid-polymer mixture, since a state point is
now specified by the densities of both components. The predicted glass
transition lines show quantitative disagreement with experiments over
the whole composition plane (cf. Fig.~1 in
\cite{reentrantGlass}). To compare calculated and measured SSFs and
DSFs, a protocol for identifying `corresponding state points' is
needed.

Secondly, the attractive interaction between two particles is always
specified directly as a potential energy in calculations. The
corresponding experimental variable is the polymer concentration in
the free volume, \cpfree. This is currently guessed at using an
uncontrolled and untested approximation based on scaled-particle
theory \cite{cptheory}, and is likely to lead to large systematic
errors in systems with high colloid volume fractions. Thirdly, the
marked and complex aging behavior of the attractive glasses
complicates the definition of a non-ergodic state for the purposes of
comparing with MCT. Despite these potential difficulties, however, our
data suggest that it may be worthwhile performing a series of
calculations at fixed $\phi$ and increasing attraction crossing the
re-entrant gap in between the repulsive and attractive glass
transition lines for a system of hard spheres interacting with
something like an Asakura-Oosawa potential \cite{gelMCTHiPhi}.

Finally, it is clear that attractive and repulsive glasses show
qualitative distinct aging behavior. Classical MCT does not predict
aging, but it is a generic feature of experimental glasses of all
kinds \cite{generalAging}. A number of theoretical approaches are
emerging, and simulation is a valuable tool. It is probable that
further study of this phenomenon in our model colloid-polymer mixture
should throw significant light on this intriguing (and generic)
phenomenon \cite{SciortinoAging}.

\acknowledgments{KNP is funded by an UK ORS award and the University
of Edinburgh. Partial financial support for experimental equipment
came from EPSRC grant GR/M92560. We thank Matthias Fuchs, Michael
Cates, Antonio Puertas, Johan Bergenholtz and Francesco Sciortino for
illuminating discussions during various stages of this work, and
Michael Cates also for commenting on the manuscript.}

\bibliography{papers}

\pagebreak

\clearpage %get around buggy figure display

\clearpage %get around buggy figure display

\end{document}